\documentclass[usenatbib]{mn2e}

\usepackage{upgreek}
\usepackage{amssymb,amsmath}
\usepackage{graphicx}
\usepackage{multirow}
\usepackage{subcaption}
\usepackage[authoryear]{natbib}
\usepackage{threeparttable}

\def\cite{\citep}

\newcommand{\Msun}{\ensuremath{\mathrm{M}_\odot}}

\title[FRB 110214]{A fast radio burst with a low dispersion measure}
\makeatletter
\author[Petroff et al.]{E.~Petroff$^{1,2,3}$\thanks{e-mail: e.b.petroff@uva.nl}\thanks{Veni Fellow}, L.~C.~Oostrum$^{1,2}$, 
B.~W.~Stappers$^{4}$,
M.~Bailes$^{2,5,6}$,
E.~D.~Barr$^{7}$,
\newauthor
S.~Bates$^{4}$,
S.~Bhandari$^{8}$,
N.~D.~R.~Bhat$^{5,9}$,
M.~Burgay$^{10}$,
S.~Burke-Spolaor$^{11,12}$,
\newauthor
A.~D.~Cameron$^{3}$,
D.~J.~Champion$^{7}$,
R.~P.~Eatough$^{7}$,
C.~M.~L.~Flynn$^{2}$,
A.~Jameson$^{2,5}$,
\newauthor
S.~Johnston$^{8}$,
E.~F.~Keane$^{13}$,
M.~J.~Keith$^{4}$,
M.~Kramer$^{4,7}$,
L.~Levin$^{4}$,
V.~Morello$^{4}$,
\newauthor
C.~Ng$^{14}$,
A.~Possenti$^{10,15}$,
V.~Ravi$^{16}$,
W.~van~Straten$^{17}$,
D.~Thornton$^{4}$,
C.~Tiburzi$^{7}$
\\
$^1$Anton Pannekoek Institute for Astronomy, University of Amsterdam, P.O. Box 94249, 1090 GE Amsterdam, The Netherlands. \\
$^2$ASTRON Netherlands Institute for Radio Astronomy, Oude Hoogeveensedijk 4, 7991PD Dwingeloo, The Netherlands. \\
$^3$Centre for Astrophysics and Supercomputing, Swinburne University of Technology, Mail H30, PO Box 218, VIC 3122, Australia. \\
$^4$Jodrell Bank Center for Astrophysics, University of Manchester, Alan Turing Building, Oxford Road, Manchester M13 9PL, UK.\\
$^5$ARC Center of Excellence for All-Sky Astronomy (CAASTRO), Swinburne University of Technology, Mail H30, PO Box 218,\\
$^{}$ VIC 3122, Australia. \\
$^6$ARC Center of Excellence for Gravitational Wave Discovery (OzGrav), Swinburne University of Technology, Mail H11, PO Box 218, \\
$^{}$ VIC 3122, Australia. \\
$^7$Max-Planck-Institut f{\"u}r Radioastronomie, Auf dem H{\"u}gel 69, D-53121 Bonn, Germany. \\
$^8$CSIRO Astronomy \& Space Science, Australia Telescope National Facility, P.O. Box 76, Epping, NSW 1710, Australia. \\
$^9$International Centre for Radio Astronomy Research, Curtin University, Bentley, WA 6102, Australia. \\
$^{10}$INAF - Osservatorio Astronomico di Cagliari, Via della Scienza 5, I-09047 Selargius (CA), Italy. \\
$^{11}$Department of Physics and Astronomy, West Virginia University, PO Box 6315, Morgantown, WV 26506, USA. \\
$^{12}$Center for Gravitational Waves and Cosmology, West Virginia University, Chestnut Ridge Research Building, Morgantown, WV 26505, USA. \\
$^{13}$ SKA Organisation, Jodrell Bank Observatory, Cheshire, SK11 9DL, UK. \\
$^{14}$Department of Physics and Astronomy, University of British Columbia, 6224 Agricultural Road, Vancouver, BC V6T 1Z1, Canada. \\
$^{15}$Università degli Studi di Cagliari, Dep. of Physics, S.P. Monserrato-Sestu Km 0,700, I-09042 Monserrato, Italy. \\
$^{16}$Cahill Center for Astronomy and Astrophysics, MC 249-17, California Institute of Technology, Pasadena, CA 91125, USA. \\
$^{17}$Institute for Radio Astronomy \& Space Research, Auckland University of Technology, Private Bag 92006, Auckland 1142, New Zealand.
}

\makeatother

\begin{document}
\maketitle

\begin{abstract}
Fast radio bursts (FRBs) are millisecond pulses of radio emission of seemingly extragalactic origin. More than 50 FRBs have now been detected, with only one seen to repeat. Here we present a new FRB discovery, FRB 110214, which was detected in the high latitude portion of the High Time Resolution Universe South survey at the Parkes telescope. FRB 110214 has one of the lowest dispersion measures of any known FRB (DM = 168.9$\pm$0.5 pc cm$^{-3}$), and was detected in two beams of the Parkes multi-beam receiver. A triangulation of the burst origin on the sky identified three possible regions in the beam pattern where it may have originated, all in sidelobes of the primary detection beam. Depending on the true location of the burst the intrinsic fluence is estimated to fall in the range of 50 $-$ 2000 Jy ms, making FRB 110214 one of the highest-fluence FRBs detected with the Parkes telescope. No repeating pulses were seen in almost 100 hours of follow-up observations with the Parkes telescope down to a limiting fluence of 0.3 Jy ms for a 2-ms pulse. Similar low-DM, ultra-bright FRBs may be detected in telescope sidelobes in the future, making careful modeling of multi-beam instrument beam patterns of utmost importance for upcoming FRB surveys.
\end{abstract}


\section{Introduction}

Fast radio bursts (FRBs) are observed as bright, millisecond radio transients of unknown origin \cite[e.g.][]{Lorimer2007,Thornton2013}. FRBs are characterized by a high dispersion measure (DM) relative to the expected contribution due to the Galaxy, corresponding to a large electron column density along the line of sight. The entire population of more than 50 FRBs observed to-date\footnote{All published FRBs are available on the FRB Catalogue; \texttt{http://www.frbcat.org}}\citep{FRBCAT} are believed to be extragalactic in origin. However, only one FRB source, FRB 121102, has been definitively localized to a host galaxy -- a dwarf galaxy at $z = 0.19273(8)$ \citep{Spitler2014,Chatterjee2017,Tendulkar2017}.

Due to their short durations ($\lesssim$ 50 ms), high flux densities ($\gtrsim$ 1 Jy), and high inferred brightness temperatures ($\gtrsim 10^{36}$ K), progenitor models involving beamed emission from compact objects are often invoked to explain FRBs. Favoured models include young, millisecond magnetars in dense progenitor environments \citep{Metzger2017}, young pulsars in nearby galaxies \citep{Connor2016SNR,CordesWasserman}, collapses of neutron stars to black holes \citep{Falcke2014}, binary neutron star mergers \citep{Totani2013}, and energetic magnetars orbiting black holes \citep{Michilli2018}. However, current observations are insufficient to trace FRBs back to any of these progenitor scenarios with confidence. Ultimately, more well-localized sources with precise measurements of intrinsic flux (i.e. not convolved with an uncertain location in a telescope beam) and distance are needed to constrain theoretical models.

\citet{Shannon2018} and \citet{MacquartASKAP2018} have recently increased the FRB population by 22 sources from a large sample detected at the Australian Square Kilometre Array Pathfinder (ASKAP). These bursts, detected in a fly's eye survey occupy a lower DM and higher fluence range than the sample from more sensitive telescopes with DMs from 114 pc cm$^{-3}$ to 991 pc cm$^{-3}$ and measured fluences from 34 -- 420 Jy ms. If the FRB source count distribution is steep, these ultrabright events are expected to occur at a lower rate than the lower fluence events typically detected by telescopes such as Parkes. 

Here we present a new FRB detected with the Parkes 64-m telescope in the High Time Resolution Universe (HTRU) South survey in 2011, FRB 110214. This FRB has one of the lowest measured FRB DMs to-date and was detected in the sidelobes of two outer beams of the Parkes multibeam receiver, implying a high intrinsic peak flux density. In \S~\ref{sec:obsSystems} we describe the observing system, in \S~\ref{sec:FRB} we present the burst properties, in \S~\ref{sec:localization} we detail our efforts to localize FRB 110214 within the Parkes beam pattern, in \S~\ref{sec:followup} we present results of follow-up at the possible locations of the FRB including searches for repeating pulses (\S~\ref{sec:repeats}) and attempts to identify a host galaxy (\S~\ref{sec:host}), and in \S~\ref{sec:discussion} we summarize these results and how they relate to the broader population of FRBs.

\section{Observations}
\subsection{Setup}\label{sec:obsSystems}

The discovery observations of FRB 110214 were part of the High Time Resolution Universe (HTRU) South survey conducted at the Parkes radio telescope in New South Wales, Australia \citep{Keith2010} using the Parkes multibeam receiver \citep[hereafter MB,][]{multibeam}. The MB has 13 circular feed horns, each of which forms an elliptical beam on the sky with a full-width half-maximum of approximately 14$\farcm$4. For the HTRU survey, each beam has a separate data stream through the Berkeley Parkes Swinburne Recorder (BPSR) which records 2-bit data to disk in the form of 1024 frequency channels across 400 MHz of bandwidth from 1.182 $-$ 1.582 GHz with 64$\upmu$s time sampling \citep{Keith2010}. Only 340 MHz of the total bandwidth is used as the top 60 MHz of the band is highly contaminated with RFI from satellites. 

The HTRU South survey data were collected between 2008 and 2014 and consisted of three survey regions at low, intermediate, and high Galactic latitudes with integration times of 4300s, 540s, and 270s, respectively. The HTRU high latitude data were partially processed by \citet{Thornton2013} leading to the discovery of FRBs 110220, 110626, 110723, and 120127. A full re-processing of the high latitude survey was done to search for FRBs using the \textsc{heimdall} single pulse search software\footnote{https://sourceforge.net/projects/heimdall-astro/} for events that match the criteria of an FRB outlined in previous publications \citep{FRB140514,Champion2016,Bhandari2018}. 

In this processing all four FRBs detected by \citet{Thornton2013} were recovered as well as six others. Five new detections were reported in \citet{Champion2016}: FRBs 090625, 121002, 130626, 130628, and 130729. A small fraction of the high latitude data, approximately 0.5\%, were not processed at the time of the \citeauthor{Champion2016} paper due to processing failures on the gSTAR supercomputer. In reprocessing these failed jobs, a new FRB was discovered. We describe the sixth detection in the following section.

\subsection{FRB 110214}\label{sec:FRB}

The fast radio burst FRB~110214 was recorded at the Parkes telescope during an observation of the high latitude portion of the HTRU South survey. The burst occurred at 2011-02-14 07:14:10.353 UTC at a reference frequency of 1.382~GHz, the middle of the observing band. In an initial search with \textsc{heimdall}, the burst was found in only a single beam (Beam 2) with S/N$_\mathrm{beam2}$~=~13, $\Delta t$~=~1.9(9) ms, and DM~=~168.8(5) pc cm$^{-3}$, one of the lowest DMs for an FRB reported thus far. The beam was centered at RA~=~01:21:17 DEC~=~$-$49:47:11 corresponding to a Galactic latitude and longitude ($\ell$, $b$) = (290.7$^\circ$, --66.6$^\circ$). For a full description of the burst properties, see Table~\ref{tab:frb}. 

\begin{figure}
\centering
\begin{subfigure}[b]{0.9\columnwidth}
\includegraphics[trim={6cm 0.5cm 8cm 1cm},clip,width=0.9\columnwidth]{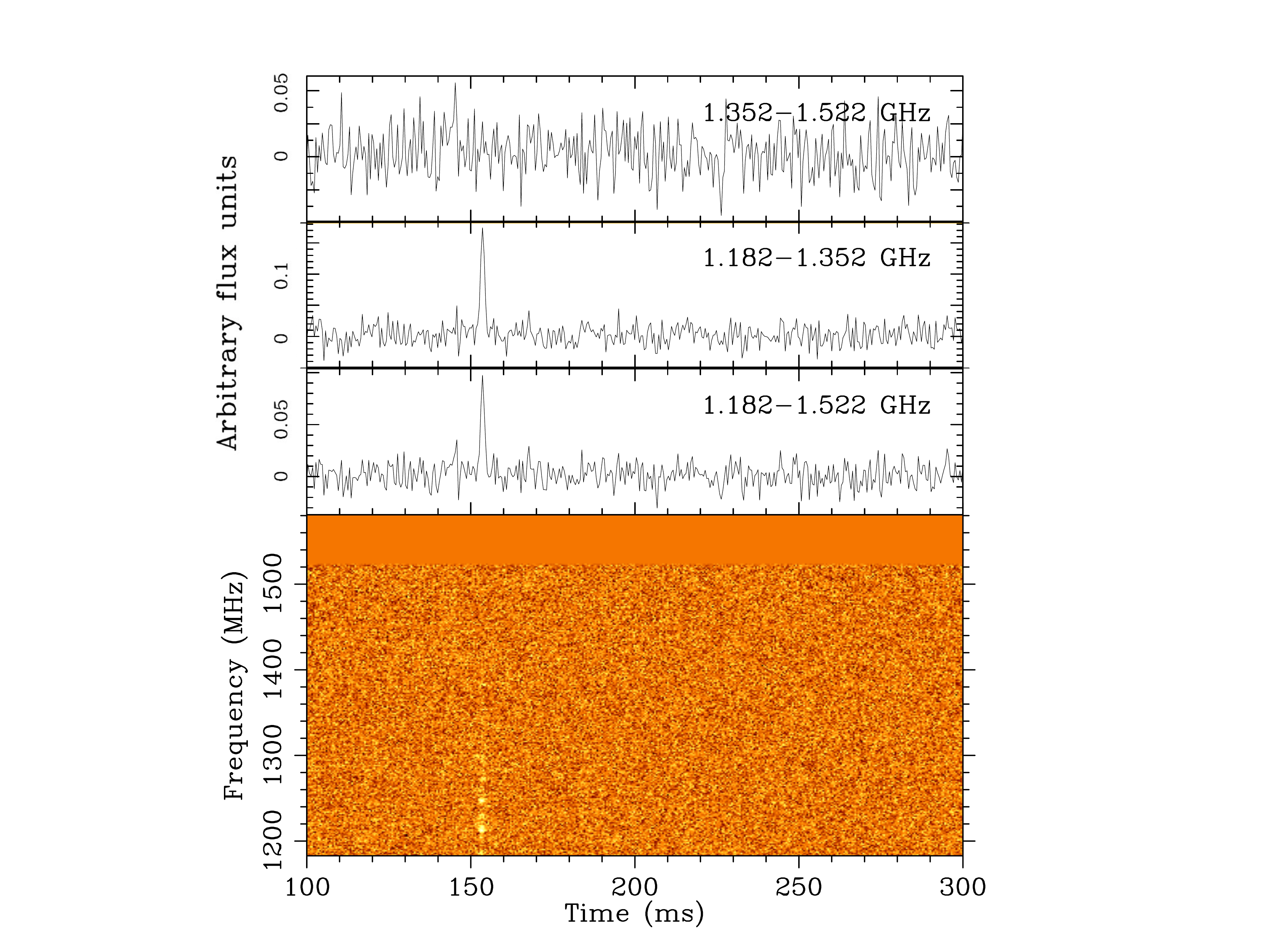}
\caption{Beam 2.}\label{fig:beam2}
\end{subfigure}

\begin{subfigure}[b]{0.9\columnwidth}
\includegraphics[trim={6cm 0.5cm 8cm 1cm},clip,width=0.9\columnwidth]{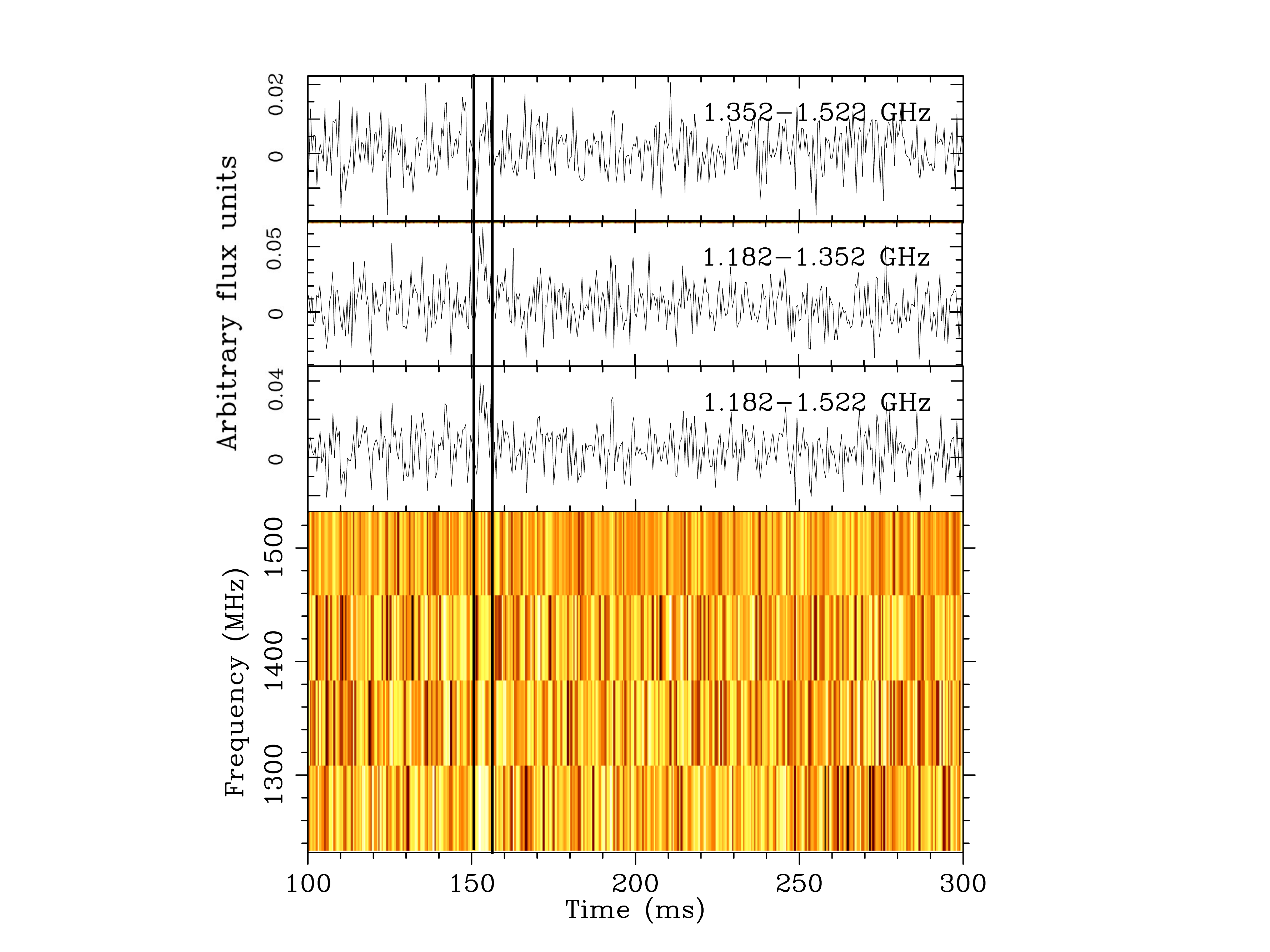}
\caption{Beam 8.}\label{fig:beam8}
\end{subfigure}
\caption{The dynamic spectra of FRB 110214 detected in (a) Beam 2 and (b) Beam 8 of the Parkes multibeam receiver. The effects of dispersion have been removed and the integrated timeseries are shown for each beam over the top half (top), the bottom half (middle), and the entire range (bottom) of the  bandwidth. The frequency channels between 1522 $-$ 1582 MHz have been masked in both beams due to persistent RFI. For Beam 8, the frequency time spectrum has been integrated into four frequency channels and the pulse range is bordered by vertical lines to guide the eye. }\label{fig:waterfall}
\end{figure}

Despite its proximity in time to other HTRU detections, having occurred only 6 days before FRB~110220, FRB~110214 was missed in previous searches. We attribute this to human error and the novelty of the field of FRBs at the time of the first search of the data. At the time these data were being searched by \citeauthor{Thornton2013} the observational characteristics of FRBs were poorly classified and a low S/N, low DM candidate present in only half of the observing band (see Figure~\ref{fig:waterfall}) may have been rejected as spurious. The DM excess of FRB~110214 is still high relative to the expected DM contribution of the Galaxy along the line of sight: 31 pc cm$^{-3}$ and 21 pc cm$^{-3}$ according to the NE2001 and YMW16 models, respectively \citep{NE2001,YMW}. However, the burst was ultimately discovered in the reprocessing of the entire survey undertaken for \citet{Champion2016} at which point it was investigated further. In deeper searches of the other beams of the MB, FRB~110214 was also found to be weakly detected in Beam 8, with S/N$_\mathrm{beam8}=$ 6 (see Figure~\ref{fig:waterfall}) and was not detected above a threshold of  S/N $\geq$ 5 in any other beams. 

The spectral index of the burst in both detection beams is negative, with significantly more signal in the lower half of the band. Integrating over the bottom 50\% of the Parkes bandwidth, from 1.182 $-$ 1.352 GHz, results in a higher significance detection in both beams with S/N$_\mathrm{beam2\_lower}$ = 17 and S/N$_\mathrm{beam8\_lower}$ = 7 (see Table~\ref{tab:SNRs}) and non-detections in all other beams. The data for this observation are not bandpass corrected; BPSR sets the levels for an observation using the first ten seconds of data. However, the BPSR bandpass shapes are consistent across many observations and close to flat. 

While the pulse may have an intrinsically negative spectral index either due to its emission process or Galactic scintillation, the detection of the burst in multiple beams of the receiver and stronger detections at lower frequencies leads to the conclusion that the source location might be far off-axis relative to both beams ($>$7\arcmin). Given the significantly reduced sensitivity of the telescope at off-axis positions, the intrinsic flux density of FRB~110214 must be high.

\begin{table}
\begin{centering}
\caption{Observed and derived properties of FRB 110214. The ranges for the peak flux and fluence of the burst are given here as derived properties due to the highly off-axis burst location and are based on the estimated origin of the FRB in the MB beam pattern; see text for further details. Derived values for redshift and distance are based on the estimated relation between DM and redshift of $z \sim \mathrm{DM}_\mathrm{excess} / 1000$ \citep{Ioka2003}.}\label{tab:frb}
\begin{tabular}{ll}
\hline
\hline
\multicolumn{2}{c}{Observed Properties} \\
Event date UTC & 2011 February 14 \\
Event time UTC, $\nu_\mathrm{1.382~GHz}$ & 07:14:10.353 \\ 
Event time, $\nu_\infty$ & 07:14:09.986 \\
Beam 2 RA & 01:21:17 \\
\hphantom{Beam 2} DEC & $-$49:47:11 \\
Beam 8 RA & 01:19:07\\
\hphantom{Beam 8} DEC & $-$49:26:40 \\
Beam 2 ($\ell$,$b$) & (290.7$^{\circ}$, $-$66.6$^{\circ}$)\\
Beam full-width half-maximum & 14.4$'$ \\
DM$_\mathrm{FRB}$ (pc cm$^{-3}$) & 168.8(5) \\
Detection S/N$_\mathrm{beam2}$ & 13(1) \\ 
Detection S/N$_\mathrm{beam8}$ & 6(1) \\
Observed width, $\Delta t$ (ms) & 1.9(9) \\ 
\hline
\multicolumn{2}{c}{Derived Properties} \\
Peak flux density, $S_{\nu,\mathrm{1200MHz}}$ (Jy) & 27 -- 1055 \\
Fluence, $\mathcal{F}$ (Jy ms) & 54 -- 2057 \\
DM$_\mathrm{MW, NE2001}$ (pc cm$^{-3}$) & 31.1 \\
DM$_\mathrm{MW, YMW16}$ (pc cm$^{-3}$) & 21.0 \\
Redshift, $z$ & $<$0.14\\
Comoving distance (Mpc) & 462 \\ 
Luminosity distance (Mpc) & 513 \\ 
\end{tabular}
\end{centering}
\end{table}

\begin{table}
\caption{The signal-to-noise ratio (S/N) of FRB 110214 in the primary (Beam 2) and secondary (Beam 8) detection beams in different sub-bands. The burst was not detected in the top sub-band of either beam.}\label{tab:SNRs}
\begin{tabular}{lccc}
\hline
\hline
 & \multicolumn{3}{c}{S/N} \\
 & 1.352$-$1.522 GHz & 1.182$-$1.352 GHz & Full band \\
\hline
Beam 2 & $<5$ & 17.0 & 13.4 \\
Beam 8 & $<5$ & 7.0 & 6.0 \\
\hline
\end{tabular}
\end{table}

\section{Localization of FRB 110214}\label{sec:localization}

To estimate the location of FRB~110214 in the beam pattern of the MB, the beam model developed for FRB~150807 by \citet{FRB150807} was modified to reflect the case of this particular burst. Briefly, the model consists of radiation patterns for each individual beam of the MB accounting for the geometry of the dish and receiver, blockage from the focus cabin, and edge tapering for each beam calculated at each point on a 1000x1000 pixel rectangular grid covering an area of 3 deg$^2$ centered on the central beam. \citeauthor{FRB150807} found this model closely approximated the observed response of the central, inner, and outer beams of the MB for bright Galactic pulsars. While any analytic model may be insufficient for the purposes of precisely pinpointing a location of the FRB on the sky, in the case of the MB it is the best approximation possible since the actual beam pattern is not fully mapped with real measurements. Nonetheless, it can provide useful information about the high probability region(s) where the burst may have originated and how bright it may have been intrinsically. 

Due to the non-detection of FRB~110214 at higher frequencies (see Table~\ref{tab:SNRs}) only the lower half of the BPSR frequency bandwidth was used in the localization analysis. The beam pattern model from \citeauthor{FRB150807} was modified to only model the MB response in the range 1.182$-$1.352 GHz. Using a method similar to that described in \citet{Obrocka2015}, the beam pattern model was searched for regions where the ratio between the signal strength in beams 2 and 8 matched that of the detected pulse in the Parkes data and accounting for the different gains in each beam. In the 170 MHz frequency band used, this corresponds to a ratio $\left(\mathrm{S/N}_\mathrm{beam2\_lower}\right) / \left(\mathrm{S/N}_\mathrm{beam8\_lower}\right) = \mathrm{S/N}_\mathrm{2:8} \sim 2.5$. Assuming an error of $\pm$1 on the detection S/N of each beam provides a window of allowable ratios $2.125 \lesssim \mathrm{S/N}_\mathrm{2:8} \lesssim 2.93$ between beams 2 and 8. Further constraints are placed by the non-detection of the pulse in any other beams with a threshold of S/N $>$ 5, such that $\mathrm{S/N}_\mathrm{2:\textit{i}} < 17.0/5.0$ for all beams $i$ except Beam 2 and Beam 8.

\begin{figure*}
\centering
\includegraphics[trim={0cm 0cm 1.5cm 1.5cm},clip,width=0.6\linewidth]{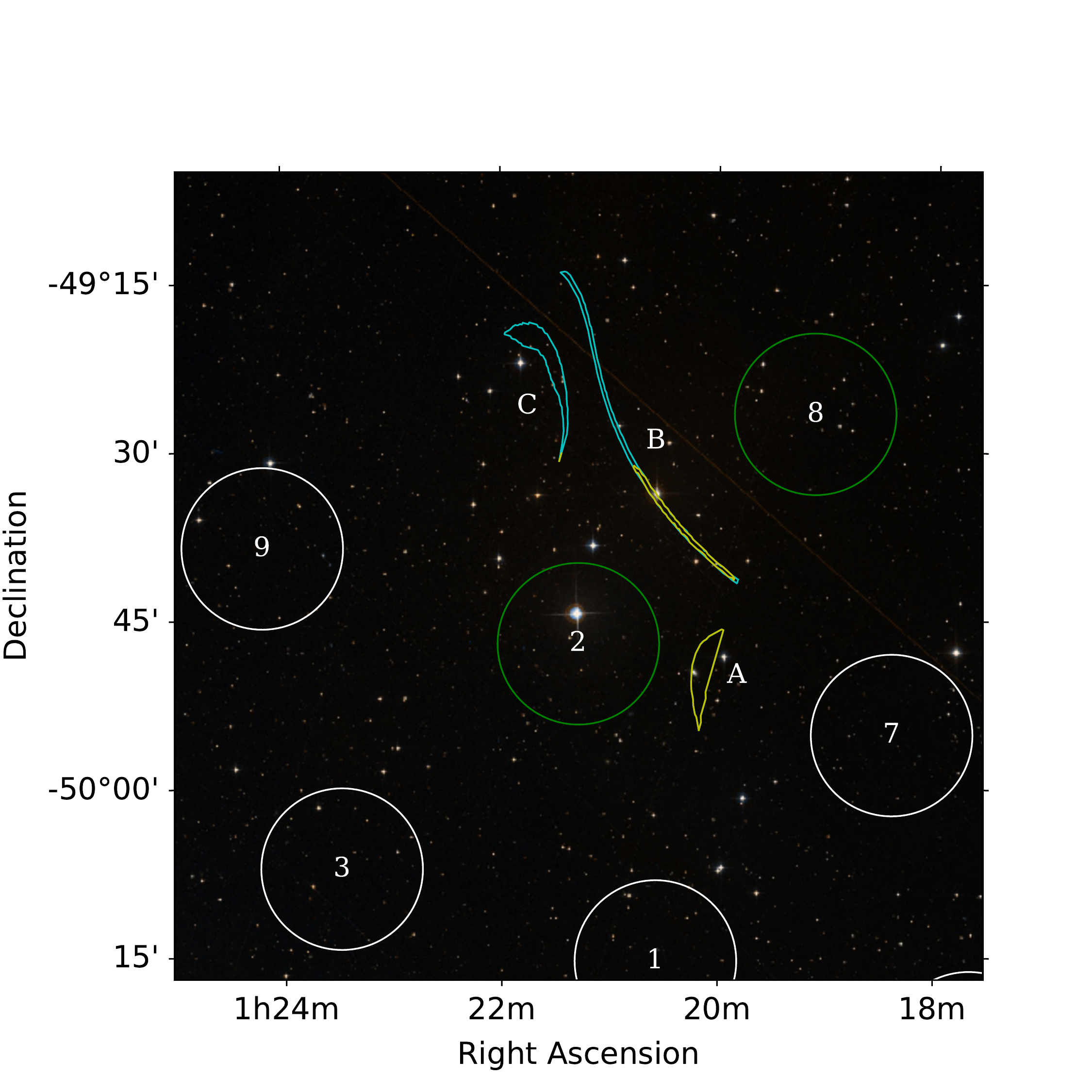}
\caption{The beams of the MB and the three 1$\sigma$ localization regions identified for FRB~110214 using the model from \citet{FRB150807} over the frequency range 1.182--1.352 GHz. The radio regions are overlaid on a sky image from the Sloan Digital Sky Survey \citep[SDSS; ][]{SDSS}. The regions (labeled A, B, and C) correspond to locations that produce the detected S/Ns of FRB 110214 in beams 2 and 8 (green), with non-detections in all other beams (white). The combined area of the 1$\sigma$ regions is roughly 52 arcmin$^2$. Assuming a Euclidean source count distribution for the FRB population, regions with a relative likelihood $\geq$10\% of the minimum flux are also shown for Regions B and C (yellow). Region A remains unchanged (see text).}\label{fig:regions}
\end{figure*}

Within these constraints, three allowed regions in the beam pattern emerge\footnote{Contours of the localization regions are provided in a supplementary file of this manuscript.}, as shown in Figure~\ref{fig:regions}. The three regions vary in distance from the primary detection beam; Region A is the closest, placing FRB~110214 in the first sidelobe of Beam 2. Region B is elongated away from the detection beams and lies along a line of constant S/N$_\mathrm{2:8}$ in the outer sidelobes. Region C is approximately 30$\arcmin$ away from the primary detection beam and would place FRB 110214 in an outer sidelobe of the MB beams. 

For each region, we estimate the intrinsic flux density and fluence of FRB 110214 required to produce the observed signal in the data. 
We calculate the intrinsic peak flux density of FRB~110214 if it were located in each region. In each case we calculate an average and median peak flux density for each region, as estimated flux density increases rapidly with distance from the center of the primary beam and can significantly affect the average for an elongated region. We calculate a fluence $\mathcal{F} = S \times \Delta t$ where we use the observed pulse width $\Delta t$ = 1.95 ms since no substantial broadening of the pulse is expected in the case of an off-axis detection. The estimated flux densities and fluences for these regions are summarized in Table~\ref{tab:RegionFlux}.

\subsection{Region A}

Region A occupies an area of $\sim$10 arcmin$^2$ (1$\sigma$) centered at RA 01:20:09 Dec --49:49:37 (11$\arcmin$ from Beam 2 center). If FRB~110214 originated in or near this region we estimate an intrinsic peak flux density of $S_\mathrm{A, avg} \approx$ 29~Jy or $S_\mathrm{A, med} \approx$ 28~Jy. These two estimates are close due to the compactness of the region and its proximity relative to the primary detection beam. These values correspond to estimated FRB fluences of $\mathcal{F}_\mathrm{A, avg} \approx$ 55 Jy ms and $\mathcal{F}_\mathrm{A, med} \approx$ 54 Jy ms. The full range of fluence values in Region A is 34 -- 95 Jy ms.

\subsection{Region B}

Region B is narrow and elongated with a total area of $\sim$14 arcmin$^{2}$ (1$\sigma$) extending from RA 01:19:48 Dec --49:41:31 at its closest to the primary beam (15$\arcmin$ from Beam 2 center) to RA 01:21:25 Dec --49:14:13 at its furthest (33$\arcmin$ from Beam 2 center). In this region the estimated peak flux density for FRB~110214 is $S_\mathrm{B, avg} \approx$ 504 Jy and $S_\mathrm{B, med} \approx$ 110 Jy. These values, in turn, correspond to estimated fluences for FRB 110214 of $\mathcal{F}_\mathrm{B, avg} \approx$ 982 Jy ms and $\mathcal{F}_\mathrm{B, med} \approx$ 215 Jy ms. The full range of fluence values in Region B is 72 -- 8472 Jy ms.

\subsection{Region C}

Region C has a large area of $\sim$28 arcmin$^{2}$ (1$\sigma$). While also highly elongated, the region is approximately centered at RA 01:21:36 Dec --49:17:34 (30$\arcmin$ from Beam 2 center). An FRB with the observed parameters of FRB~110214 originating from this region would have an intrinsic flux $S_\mathrm{C, avg} \approx$ 899 Jy or $S_\mathrm{C, med} \approx$ 1055 Jy. This corresponds to a fluence of $\mathcal{F}_\mathrm{C, avg} \approx$ 1754 Jy ms or $\mathcal{F}_\mathrm{C, med} \approx$ 2057 Jy ms. The full range of fluence values in Region C is 127 -- 2996 Jy ms.

\subsection{Spectral properties}

The presence of three regions here is not particularly surprising. The detection significance of FRB 110214 is much lower than in the case of of other multi-beam FRBs such as FRB~010724 and FRB~150807 \citep{Lorimer2007,FRB150807}, making the triangulation of the burst location on the sky more challenging. In the cases of FRB 010724 and FRB 150807 an additional localization constraint could be made using multiple sub-bands. No regions were identified in the beam pattern matching the constraints outlined in Section~\ref{sec:localization} that also satisfied the condition of non-detection in the top half of the band assuming a flat spectrum. Thus the FRB itself must have a negative spectral index either due to intrinsic emission or propagation effects along the line of sight. Only weak constraints on the spectral index or localization can be derived due to the complete lack of signal at these frequencies. All calculations for the intrinsic peak flux density and fluence of FRB~110214, therefore, only consider emission over 50\% of the Parkes observing band, as this is where signal was present for analysis.

\begin{table}
\caption{Estimated intrinsic peak flux density $S$ and fluence $\mathcal{F}$ of FRB~110214 for an origin in each of the localization regions identified in the beam pattern. Flux density and fluence are calculated using data in the frequency range 1.182 -- 1.352 GHz. In each case an average (avg) and median (med) value for flux density and fluence is given.}\label{tab:RegionFlux}
\begin{tabular}{c|cc|cc}
\hline
\hline
Region & $S_\mathrm{avg}$ & $\mathcal{F}_\mathrm{avg}$ & $S_\mathrm{med}$ & $\mathcal{F}_\mathrm{med}$ \\
 & (Jy) & (Jy ms) & (Jy) & (Jy ms) \\
\hline
A & 28.5 & 55 & 27.8 & 54 \\
B & 504 & 982 & 110 & 215 \\
C & 899 & 1754 & 1055 & 2057 \\
\hline
\end{tabular}
\end{table}

\section{Follow-up observations}\label{sec:followup}

\subsection{Search for repeating pulses}\label{sec:repeats}

Given the low DM and the large implied fluence of 50 -- 2,000 Jy ms for FRB~110214, significant time was spent on follow-up to search for repeating pulses. Repeating pulses from FRB 121102 are several orders of magnitude fainter in peak flux density \citep{Spitler2016,Chatterjee2017}, but the distance to the host galaxy of the repeating FRB is almost twice that of the estimated distance to FRB 110214 \citep{Tendulkar2017}. The burst was identified in the HTRU data in February 2016. Due to the commissioning of the Effelsberg phased array feed \citep[PAF; ][]{ParkesPAF} at the time, the only available receiver at the Parkes focus was the single pixel H-OH receiver\footnote{https://www.parkes.atnf.csiro.au/observing/documentation/\newline user\_guide/pks\_ug\_3.html\#Receiver-Fleet}. The H-OH receiver was used over a 256 MHz bandwidth centered at 1.386 GHz with a beam full-width half-maximum (FWHM) of 14$\farcm$8. A total of 62.5 hours of follow-up were conducted with the H-OH receiver with 96$\upmu$s time resolution centered at the position RA 01:20:13 Dec --49:49:47, in Region A. No single pulses were found at any DM $\leq$ 5000 pc cm$^{-3}$ above S/N$>$5 over the entire bandwidth, corresponding to a flux density threshold of 0.15 Jy for a 2-ms pulse. These data were heavily affected by RFI since no multi-beam coincidence could be used for candidate rejection and we estimate that approximately 5\% of the data were ruined by interference. This estimate is derived from the total number of time samples which had to be masked due to the presence of impulsive broadband RFI across all observations. 

The early choice to focus on Region A was motivated by a limited amount of telescope time, the availability of only a single-pixel receiver, and the presence of a bright nearby galaxy in the region (Section~\ref{sec:host}). An origin in Region A would also imply the lowest intrinsic flux density of 17 $>$ $S_\mathrm{peak}$ $>$ 49 Jy, thought to be the most likely as ultra-bright FRBs such as FRB 010724 \citep[800$\pm$400 Jy;][]{ParkesFRBs}, FRB 170827 \citep[50 Jy;][]{Farah2018}, and FRB 180309 \citep[$>$20 Jy;][]{FRB180309} were not as common, nor were the high fluence ASKAP FRBs known.

In December 2016 the MB was re-installed in the Parkes focus cabin and additional follow-up efforts were undertaken. An additional 32.7 hours of follow-up were conducted with the MB with the central beam centered on Region A such that outer beams covered the majority of Regions B and C. These data were searched with \textsc{heimdall} for pulses matching the same criteria as above and no pulses were found at any DM above S/N$>$5 over the entire bandwidth, corresponding to a flux density threshold of 0.13 Jy for a 2-ms pulse.

\subsection{Identifying a host galaxy}\label{sec:host}

Each region was matched to several catalogues with the aim of identifying a possible host galaxy. No sources were found in the Chandra Source Catalog or the XMM-Newton Serendipitous Source Catalogue \citep{Chandra,XMM}. The near-infrared Vista Hemisphere Survey \citep[VHS; ][]{VHS} only covers region C and two thirds of region B, however it already identifies about 250 objects as possible galaxies in those regions. As redshifts are not available for these sources, we are unable to reduce this number by considering the maximum expected redshift of the FRB.

All regions are fully covered by the  2-micron All-Sky Survey \citep[2MASS; ][]{2MASS}. It reports one known galaxy in region A at RA 01:20:13.411 DEC --49:49:47.64. This galaxy, FRL~692, is an elliptical galaxy at a redshift of $z \approx 0.025$ \citep{FRL}. At this redshift, the IGM is expected to contribute $\sim$ 17 pc cm$^{-3}$ -- 25 pc cm$^{-3}$ to the total DM from the \citet{YMW} and \citet{Ioka2003} models, respectively. The total DM in the host galaxy (i.e. the excess from the Galaxy and the IGM) would then be 131 pc cm$^{-3}$ using YMW16 and 113 pc cm$^{-3}$ using NE2001 and the IGM model from \citet{Ioka2003}. We do not expect a significant contribution to the DM due to an interstellar component in elliptical galaxies \citep{Xu2015}; however, if the source of the FRB were embedded in an ionized progenitor region like FRB 121102, such a host contribution could be feasible.

Considering the possibility that the host of FRB 110214 is similar to that of FRB 121102, we estimate the total number of possible host galaxies assuming the host to be a dwarf galaxy at least as massive as the host galaxy of FRB 121102 \citep[$(4-7) \times 10^7 \Msun$; ][]{Tendulkar2017}. The number density of galaxies can be described by the Schechter function

\begin{equation}
\Phi (M)\,\mathrm{d}M = \phi_* \, (M/M_*)^{\alpha} \, \mathrm{e}^{M/M_*} \, \mathrm{d}M,
\end{equation}

\noindent where $\Phi(M)$ is the number density of galaxies per unit mass, $M_*$ is the characteristic mass and $\phi_*$ is a normalization constant. We consider two mass functions: 1) the stellar mass function for blue galaxies \citep[defined as $u-r\lesssim1.9$;][]{Baldry2012}, which should more closely resemble the repeater host galaxy as they have a higher specific star formation rate than red galaxies, and 2) the H{\sc i} mass function \citep{Haynes2011}, which is likely more complete for dwarf galaxies as they tend to have relatively high H{\sc i} to stellar mass ratios ($1-10$), making them easier to detect in H{\sc i}. Integrating these mass functions from $M_\mathrm{stellar} = 4 \times 10^7 \Msun$ to the maximum mass considered to be a dwarf galaxy, $10^{10} \Msun$, and allowing the H{\sc i} to stellar mass ratio to vary between 1 and 10, gives a dwarf galaxy number density of $n = (0.02 - 0.06)\,\mathrm{Mpc^{-3}}$.

\begin{figure}
\centering
\includegraphics[width=0.9\columnwidth]{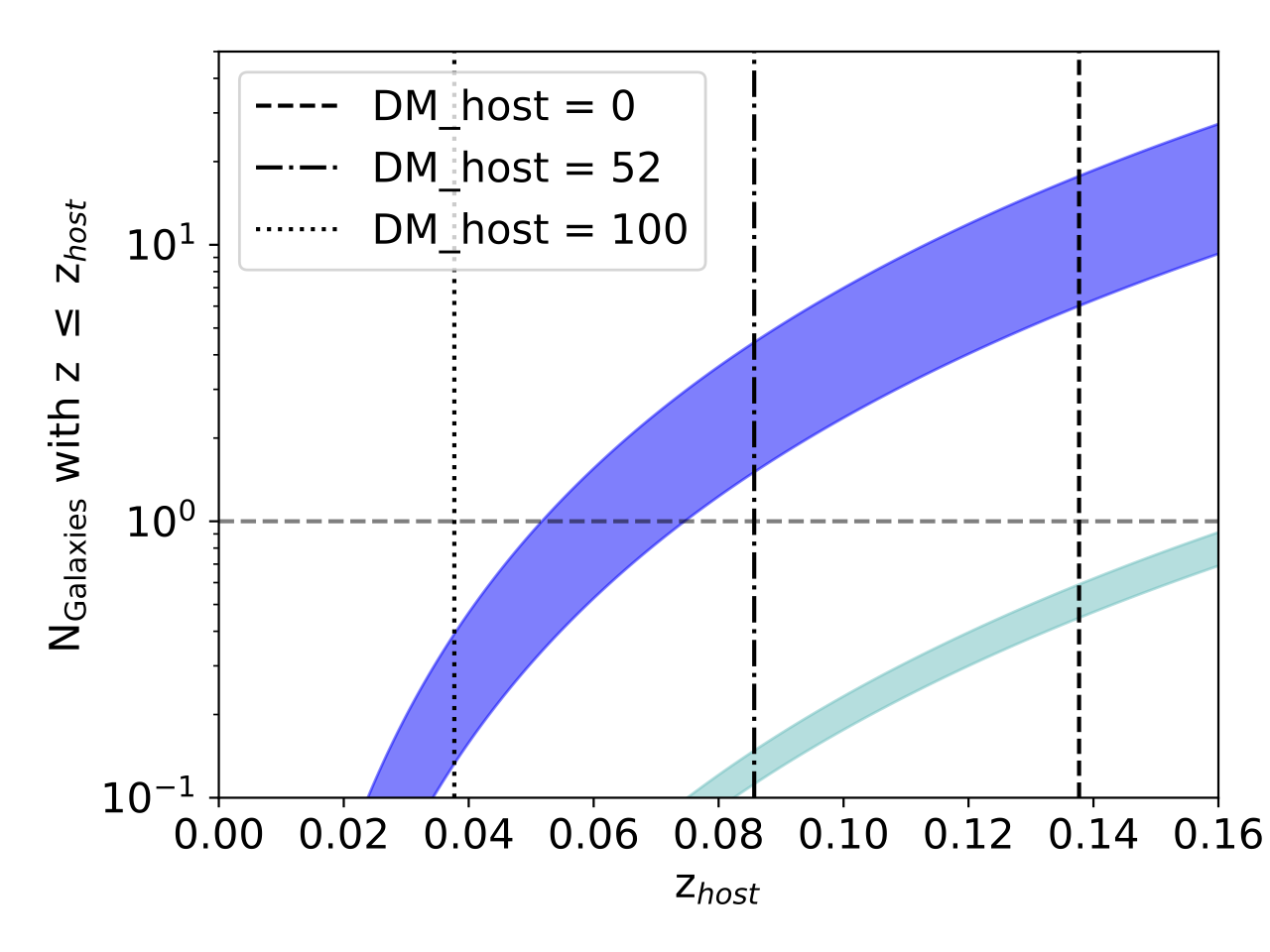}
\caption{The expected number of galaxies in the FRB 110214 localization area for a range of dwarf galaxy (dark blue), and massive galaxies (cyan) number densities. The horizontal dashed line indicates where a single galaxy is expected. The three vertical lines indicate the host galaxy redshift for a host DM contribution of $0$ pc cm$^{-3}$ (dashed), $100\,\mathrm{pc\,cm^{-3}}$ (dotted), and 52 pc cm$^{-3}$ (dot-dashed) which gives the same IGM to host DM ratio as FRB 121102.}\label{fig:ngalaxy}
\end{figure}

Assuming the mass function and galaxies do not evolve significantly between $z=0$ and $0.14$, which is reasonable given the low redshift, the total expected number of dwarf galaxies in the FRB 110214 error region is simply $n\,V_\mathrm{C}$, where $V_\mathrm{C}$ is the comoving volume, which we calculate from the redshift assuming the best-fit cosmological parameters of \citet{Planck2016}. The number of dwarf galaxies as a function of redshift is shown in Figure~\ref{fig:ngalaxy}. We also show the expected number of massive $M_\mathrm{stellar} > 10^{11} \Msun$ galaxies, based on the luminosity function of blue galaxies of \citet{faber2007}, from which we find a number density of $n = (1.5 - 2.0)\times10^{-3}\,\mathrm{Mpc^{-3}}$. For a host DM contribution of zero, between 5 and 20 dwarf galaxies are expected within the the total volume of all three regions. In contrast, no massive galaxies are expected at all. Hence, the proximity of FRL~692 is unlikely, but without more precise localization of the FRB we cannot say anything about an association with certainty.

\section{Discussion and Conclusions}\label{sec:discussion}

FRB 110214 has one of the lowest DMs of the entire FRB sample to-date. The multi-beam detection and also the fact that it was only seen in the lower half of the band indicate that the FRB occurred in the MB sidelobes. From our analysis of the best available model of the MB beam pattern we have identified three possible regions on the sky where the burst may have originated at different separations from the center of the primary beam. From these locations relative to the primary beam we have estimated the intrinsic fluence of the burst in the bottom half of the bandwidth where it would be 50 Jy ms if located in the innermost region, to a maximum of $\sim$2,000 Jy ms if located in the outermost. Even in the most conservative case, this would make FRB 110214 one of the highest fluence FRBs detected with Parkes. Despite the high implied brightness, the required energetics to produce the burst are still consistent with those of other known FRBs due to the very low inferred distance. Assuming the burst originated at the maximum estimated redshift of $z = 0.14$ the total isotropic energy required would be $E_\mathrm{FRB,\,iso} = 3.7 - 135 \times 10^{32}$ J ($\times 10^{39}$ erg) for the entire range of estimated fluences in Table~\ref{tab:RegionFlux}. Even at the high end of this range, the implied energy is still lower than that of FRB 160102 with $E_\mathrm{160102,\,iso} = 628 \times 10^{32}$ J if it originated at its maximum implied redshift $z = 2.1$. 

Not all the localization regions identified for FRB~110214 are equally likely if FRBs are isotropically distributed with a steep brightness distribution. Assuming a Euclidean distribution with a log$N$--log$S$ slope of $\alpha = -1.5$, regions where the peak flux density of FRB 110214 is lower are more probable. Based on this assumption, we calculate a fractional likelihood for all locations in each region relative to the minimum peak flux density of 17.7 Jy, which occurs in Region A. For example, considering all locations with a fractional likelihood $(S_i / S_\mathrm{min})^{-1.5} \geq 0.10$ or 10\% results in the exclusion of 90\% of Region C and 54\% of Region B. All of the locations in Region A are above this threshold. These more limited probability regions contours are shown in Figure~\ref{fig:regions} in yellow.

However, the true flux distribution of FRBs remains unknown. A shallower distribution of $\alpha = -0.6$ has been suggested by \citet{Harish2016} and early results from ASKAP suggest a steeper than Euclidean distribution of $\alpha = -2.1^{+0.6}_{-0.5}$ \citep{Shannon2018}. The latter case would further favor Region A. A larger statistical sample is needed, however, to determine the true value.

An ultra-bright FRB originating in one of the outer regions is still possible, but in either case an origin closer to the detection beam is more likely. Despite significant follow-up efforts with the Parkes telescope, FRB 110214 has not been seen to repeat; most follow-up presented here was focused on searches in Region A. The low DM (and thus small implied distance) and the high intrinsic fluence make this source an excellent candidate for further follow up. A sensitive telescope such as Parkes or MeerKAT centered on or near the true sky position should be able to detect fainter repeating pulses, even if the source pulse energy distribution were steep. In our monitoring observations, no pulses were detected above a flux density of 0.15 Jy in almost 100 hours of follow-up.

There was a bright galaxy in the innermost localization region of FRB 110214, the elliptical galaxy FRL 692 at $z \simeq 0.025$. This is a similar case to the lowest DM FRB of the ASKAP sample, FRB 171020 which was found to have one bright and potentially interesting field galaxy in the error region, ESO 601$–$G036 \citep{Mahony2018}. The error region for FRB 171020 (0.38 deg$^2$) was much larger than that of FRB 110214  (0.014 deg$^{2}$) and thus the large positional uncertainty made it similarly difficult to precisely identify a host. Ultimately, more precise localization can only be achieved through the detection of repeating pulses.

If FRB 110214 is found to repeat and can be localized through its single pulses it could provide us with one of the closest FRB host galaxies available for study. A host galaxy in the local Universe, particularly if FRB 110214 is found to reside in a dwarf galaxy like FRB 121102, would provide a rich opportunity to study the structure and composition of the host that have been limited for FRB 121102 due to low signal-to-noise and the long integration times necessary to obtain spectra. Thus, although the total available observing time for follow-up of archival FRBs is limited, we argue that FRB 110214 should be one of the top priorities for monitoring campaigns in the future. 

The localization regions presented here are not exact, as the MB beam model used is only an approximation. The sky around the regions identified here should also be monitored, such as with a phased array feed on a single dish, or with a number of overlapping or adjacent beams formed on the field with an interferometer. FRB surveys are sensitive to bright bursts over more of the sky, and high fluence events like FRB 110214 will be detectable in the telescope sidelobes. Thus, future surveys with multi-beam instruments should take great care to model and understand the beam and sidelobe patterns of their instruments in order to more accurately localize bright bursts on the sky.

\section*{Acknowledgements}

The authors thank the anonymous referee for their comments and feedback which improved the quality of the manuscript. The Parkes radio telescope is part of the Australia Telescope National Facility which is funded by the Commonwealth of Australia for operation as a National Facility managed by CSIRO. Parts of this research were conducted by the Australian Research Council Centre of Excellence for All-sky Astrophysics (CAASTRO), through project number CE110001020 and the ARC Laureate Fellowship project FL150100148. This work was performed on the gSTAR national facility at Swinburne University of Technology. gSTAR is funded by Swinburne and the Australian Government's Education Investment Fund. EP and LCO acknowledge funding from the European Research Council under the European Union’s Seventh Framework Programme (FP/2007-2013)/ERC Grant  Agreement No. 617199. SBS is supported by NSF award \#1458952. BWS acknowledges funding from the European Research Council (ERC) under the European Union’s Horizon 2020 research and innovation programme (grant agreement No 694745]). This research made use of data obtained from the Chandra Source Catalog, provided by the Chandra X-ray Center (CXC) as part of the Chandra Data Archive. This research made use of Astropy,\footnote{http://www.astropy.org} a community-developed core Python package for Astronomy \citep{astropy:2013, astropy:2018}. 

\bibliographystyle{mnras}
\bibliography{FRB110214}

\end{document}